\documentstyle[aps]{revtex}

\newcommand{\ii}{{\rm i}}
\newcommand{\de}{{\rm\,d}}
\newcommand{\e}{{\rm e}}

\newcommand{\bm}{\bbox}
\newcommand{\g}{\gamma}
\newcommand{\sig}{\sigma}
\newcommand{\eps}{\epsilon}
\newcommand{\nn}{\nonumber}
\newcommand{\ovl}{\overline}

\newcommand{\Sslash}{\rlap{/} S}

\newcommand{\nslash}{\rlap{/} n}

\begin{document}
 
\title{
\begin{flushright}
\begin{minipage}{4 cm}
\small
VUTH 01-09\\
\end{minipage}
\end{flushright}
Positivity bounds on spin-one distribution and fragmentation functions}

\author{A.~Bacchetta\footnote{Corresponding author. E-mail:{\tt
bacchett@nat.vu.nl}},  P.~J.~Mulders}

\address{
$^1$Division of Physics and Astronomy, Faculty of Science, Free University \\
De Boelelaan 1081, NL-1081 HV Amsterdam, the Netherlands\\[2mm]}

\date{draft of June 12th, 2001}
\maketitle
\begin{abstract}
We establish a connection between the distribution functions
of quarks in spin-one hadrons and the helicity matrix for 
forward scattering of antiquarks off spin-one hadrons. 
From positivity of this matrix we obtain inequality 
relations among the distribution functions. Analogous relations hold also
for fragmentation functions.
The bounds we obtained can
be used to constrain estimates of 
unknown functions, occurring in particular in semi-inclusive deep inelastic 
scattering or
$e^+e^-$ annihilation with vector mesons in the final state.
\end{abstract}

\section{Introduction}

In recent years some attention has been given to distribution functions
characterizing  spin-one targets, starting from the work of
Hoodbhoy, Jaffe and Manohar~\cite{hjm}. 
Unfortunately, the only
available spin-one target is the deuteron, which is essentially a weakly bound
system of two spin-half hadrons. In the approximation of independent
scattering off the nucleons, the specific spin-one contribution
to the deuteron structure functions is expected to be
small, except maybe for the low-$x$ region \cite{Nikolaev:1997jy,Bora:1998pi}.
New precision measurements of the deuteron 
structure functions coming from HERMES may provide an experimental test of
this expectation, as already
observed in several papers~\cite{ael,kum,et}.

Instead of pointing the attention to spin-one targets,
 it is also possible to
analyze spin-one final state hadrons in $e^+\,e^-$ annihilation or in
semi-inclusive deep inelastic scattering. This idea was first considered by
Efremov and Teryaev~\cite{Efremov:1982vs}. A systematic study was
accomplished by Ji~\cite{ji}, who singled out two new fragmentation functions,
the function $\hat{b}_1$, in analogy to the distribution 
function $b_1$, and  the time-reversal odd (T-odd) function
$\hat{h}_{\overline{1}}$. These new
fragmentation functions can be observed in the production 
of vector mesons,
{\it e.g.}\ $\rho$, $K^{\ast}$, $\phi$.
However, these functions require polarimetry on the final-state meson, which
can be done by studying the angular 
distribution of its
decay products ({\it e.g.}\ $\pi^+ \, \pi^-$ in the case of $\rho^0$ meson). In this
sense, vector meson fragmentation functions represent just a specific 
contribution to the more general 
analysis of two-particle production~\cite{rai,jjt} near the vector 
meson mass. 

In a recent work~\cite{io} we contributed to the study of spin-one 
distribution and
fragmentation functions by defining all possible functions occurring at
leading order in $1/Q$ when also partonic transverse momentum is included,
extending the work done for spin-half hadrons in~\cite{multa}. 

The study of spin-one fragmentation functions has a twofold relevance: it
offers new insights in the physics of mesons and it serves as a probe of
particular nucleonic distribution 
functions. With regards to the second issue, we point out that
for 1-particle inclusive leptoproduction of a spin-one meson, 
at leading (zeroth) order in $1/Q$ there are five chiral-odd fragmentation 
functions that can be observed in combination with the quark transversity 
distribution function~\cite{brook}, 
the measurement of which is at present attracting the attention of 
experiments such as HERMES, COMPASS and RHIC. 

In this situation, it seems to be extremely useful to gather as much
information as possible about spin-one distribution and fragmentation 
functions. As a
first step in this direction, we present here positivity bounds on them, 
following
what has been done in Ref.~\cite{noi} for the spin-half case.

\section{Bounds on transverse momentum integrated functions}

Distribution functions appear parametrization of the
 light-cone correlation function
~\cite{Soper77,Jaffe83,Manohar90}
\begin{equation} 
\Phi_{ij}(x) = \left. \int \frac{\de \xi^-}{2\pi}\ \e^{\ii p\cdot \xi}
\,\langle P,S,T\vert \overline \psi_j(0) \psi_i(\xi)
\vert P,S,T\rangle \right|_{\xi^+ = \xi_T = 0},
\label{e:dist}
\end{equation}
depending on the light-cone fraction $x = p^+/P^+$, where $P$ is
the momentum of the target hadron and $p$ the momentum of the outgoing
quark.  Since we are dealing with
spin-one hadrons the polarization state of the target must be specified using
a spin vector $S$ and a spin tensor $T$. In the target rest-frame, we can
conveniently parametrize the spin vector and tensor as
\footnote{Note that the the parametrization of $T$ is different from the one
of Ref.~\cite{io}. There is a factor $-2/3$ difference between the old
definition of $S_{LL}$ and the new, so that now 
$-1/3 \leq S_{LL} \leq 2/3$.}
\begin{eqnarray}
{\bm S}&=&\left \lgroup S_{T}^x, S_{T}^y, S_{L}\right \rgroup, 
   \label{e:vec}\\
&&\nonumber \\
{\bm T}&=&\frac{1}{2}\left \lgroup \begin{array}{ccc}
	{S_{LL}}+{S_{TT}^{xx}} 
		& {S_{TT}^{xy}}	& {S_{LT}^{x}} \\ \\
	{S_{TT}^{xy}}& 	{S_{LL}}-{S_{TT}^{xx}}
					& {S_{LT}^{y}}  \\ \\
	 {S_{LT}^{x}}	& {S_{LT}^{y}}	& -2\,{S_{LL}} 
	\end{array}\right \rgroup.
\label{e:tens}
\end{eqnarray} 

At leading twist (leading order in $1/Q$) in inclusive leptoproduction, the
relevant part of the correlator is $\Phi\gamma^+$. Its transpose 
represents the forward amplitude for antiquark-hadron scattering, 
$M=\left( \Phi \, \g^+ \right)^{T}$. 
Being the square of a scattering amplitude, the matrix $M$ in the 
parton $\otimes$ hadron spin space  is positive definite, 
\begin{eqnarray} 
M_{ij, \lambda' \lambda}(x)
& = & \left. \int \frac{\de \xi^-}{2\pi\sqrt{2}}\ \e^{\ii p\cdot \xi}
\,\langle P,\lambda'\vert \psi^\dagger_{+i}(0)\, \psi_{+j}(\xi)
\vert P,\lambda\rangle \right|_{\xi^+ = \xi_T = 0}
\nn \\
& = & \frac{1}{\sqrt{2}}\sum_n
\langle P_n\vert \psi_{+i}\vert P,\lambda'\rangle^\ast
\,\langle P_n\vert \psi_{+j}\vert P,\lambda\rangle
\,\delta\left(P_n^+ - (1-x)P^+\right) ,
\label{dens}
\end{eqnarray} 
where 
$\psi_+ \equiv {\cal P}_+\psi = \frac{1}{2}\gamma^-\gamma^+\psi$ is the
good component of the quark field~\cite{KS70} and where 
$\lambda, \, \lambda'=1,0,-1$.

For spin-one hadrons, $\Phi(x) \gamma^+$ is 
parametrized in terms of five distribution functions
\begin{equation}  
\Phi (x)\g^+ =\Bigl\{
		f_1(x) +
		g_{1}(x)\,S_{L}\,\g_5 +
		h_{1}(x)\,\g_5 \Sslash_T 
		\mbox{}+
		f_{1LL}(x)\,S_{LL} +
		\ii\,h_{1LT}(x)\,\Sslash_{LT}\Bigr\}
		{\cal P}_+.		 
\label{e:phi}
\end{equation} 
The last distribution function is constrained to be zero by time-reversal
invariance. Nevertheless, we choose to keep it because in the
analysis of fragmentation functions the analogous contribution, $H_{1LT}(z)$,
cannot be constrained in the same way. 
To follow a more systematic naming of the functions when we will include
transverse momenta, we felt the need to
change the names of the transverse momentum independent functions. We are
aware that changing notations often poses some undesired difficulties, but we
believe in the convenience of using a notation that harmoniously connects
transverse momentum dependent to transverse momentum integrated functions.  
With the parametrizations in Eq.~(\ref{e:tens}) and Eq.~(\ref{e:phi}),
however,  
our function $f_{1LL}$ is {\it precisely the same} as the function 
$b_1$ of Hoodbhoy, Jaffe and Manohar \cite{hjm}.
 
To construct the matrix $M$ for the leading order correlation function, only
two basis states in  Dirac space are relevant, 
corresponding to good components of the right and the left handed
partons. This is particularly transparent when writing Eq.~(\ref{e:phi})
in chiral representation. Therefore,  we can effectively reduce the 
four-dimensional 
Dirac space to the $2 \times 2$  good parton chirality space. 
In other words, 
only the scattering of good antiquarks off hadrons gives a non vanishing
contribution to leading twist.

The leading-twist scattering matrix takes then the form 
\begin{eqnarray}
\lefteqn{M_{ij}(x; S,T)=} \nn \\
&& \left\lgroup \begin{array}{cc}
f_1(x) + g_1(x)\,S_L +f_{1LL}(x)\,S_{LL} & 
	{h_1}(x)\,(S_T^x+\ii \,S_T^y)
	+\ii\, {h_{1LT}(x)}\,(S_{LT}^x+\ii \,S_{LT}^y) \\
& \\
	{h_1}(x)\,(S_T^x-\ii \,S_T^y)
	-\ii\, {h_{1LT}(x)}\,(S_{LT}^x-\ii \,S_{LT}^y)  &
f_1(x) - g_1(x)\,S_L +f_{1LL}(x)\,S_{LL}.
\end{array}\right\rgroup \nn \\
\label{e:Mij}
\end{eqnarray}  
The result contains the dependence on the hadron spin vector and tensor. 
We aim at obtaining a $6 \times 6$  matrix encompassing
 also the  hadron spin space, with no dependence on the spin vector and
tensor. To this goal, we need to use the hadron spin density matrix, 
$\rho(S,T)$, and invert the relation
\begin{equation} 
M_{ij}(x; S,T) = \rho_{\lambda' \lambda}(S,T)\;
		M_{i \lambda,\,j \lambda'}(x).
\label{e:mandrho}
\end{equation}

The spin density matrix for spin-one hadrons can be decomposed on a Cartesian
basis of $3\times 3$ matrices, using the (rank-one) spin vector, $S^i$,
and the rank-two spin tensor, $T^{ij}$,
\begin{equation} 
{\bm \rho}=\frac{1}{3}\left({\bm 1} + \frac{3}{2} S^i {\bm\Sigma}^i 
		+ 3\, T^{ij} {\bm\Sigma}^{ij}\right),
\label{e:density}
\end{equation} 
where the matrices ${\bm\Sigma}^i$ are the generalization of the Pauli
matrices to the three-dimensional case and
where we chose
\begin{equation} 
{\bm\Sigma}^{ij}=\frac{1}{2}\left({\bm\Sigma}^i {\bm\Sigma}^j 
	+{\bm\Sigma}^j {\bm\Sigma}^i\right)
		-\frac{2}{3}{\bf 1} \; \delta^{ij}.
\end{equation} 

Inverting Eq.~(\ref{e:mandrho}), we can eventually
reconstruct the $6 \times 6$ leading-twist scattering matrix 
\begin{eqnarray} 
\lefteqn{M_{i \lambda,\,j \lambda'}(x)=}&&\nn\\
&&\hspace{-5mm}{\left\lgroup \begin{array}{cccccc}
{f_1} + {g_1}-\frac{{f_{1LL}}}{3} & 0 & 0 & 0 & 
	\sqrt{2}\,({h_1}+\ii \,{h_{1LT}})& 0 \\
& \\
0 & {f_1} +\frac{2\,f_{1LL}}{3} & 0 & 0 & 0 &\sqrt{2}\,({h_1}-\ii\, {h_{1LT}}) \\
& \\
0 & 0 &{f_1} - {g_1}-\frac{{f_{1LL}}}{3} & 0 & 0 & 0 \\
& \\
0 & 0 & 0 & {f_1} - {g_1}-\frac{{f_{1LL}}}{3}& 0 & 0 \\
& \\
\sqrt{2}\,({h_1}-\ii\, {h_{1LT}}) & 0 & 0 & 0 &{f_1} +\frac{2\,f_{1LL}}{3} & 0 \\
& \\
0 &\sqrt{2}\,({h_1}+\ii\, {h_{1LT}}) & 0 & 0 & 0 & 
		{f_1} + {g_1}-\frac{{f_{1LL}}}{3}
\end{array}\right\rgroup}, \nn \\
\label{e:emme}
\end{eqnarray}  
where all the functions on the right-hand side depend only on $x$. 

From the positivity of the diagonal elements we obtain the bounds:
\begin{eqnarray} 
f_1 (x) &\geq&  0 \\
-\frac{3}{2}\,f_1(x) &\leq& f_{1LL}(x) \;\leq\; 3\,f_1 (x) \\
|g_1 (x)|&\leq& f_1 (x)-\frac{1}{3}\,f_{1LL}(x)\;\leq\;\frac{3}{2}\,f_1(x),
\end{eqnarray} 
while positivity of 2-dimensional minors gives the
bound~\cite{Soffer73}  
\begin{equation} 
\left[h_1(x)\right]^2 + \left[h_{1LT}(x)\right]^2
	\leq \frac{1}{2}\Big[f_1(x)+\frac{2}{3}\,f_{1LL}(x)\Big] 
		\left[f_1(x)+g_1(x)-\frac{f_{1LL}(x)}{3}\right],
\label{e:soffer}
\end{equation} 
This bound is a generalization of the Soffer bound and must be fulfilled by
any spin-1 target. When $h_{1LT}=0$ due to time-reversal invariance, 
the bound can be reduced to
\begin{equation} 
|h_1(x)|\leq \sqrt{\frac{1}{2}\Big[f_1(x)+\frac{2}{3}\,f_{1LL}(x)\Big] 
		\left[f_1(x)+g_1(x)-\frac{f_{1LL}(x)}{3}\right]}.
\end{equation}

An analogous calculation can be performed for fragmentation 
functions, which describe the hadronization process of a parton into the 
final detected hadron. In this case the transverse momentum independent 
correlator is~\cite{CS82}
\begin{equation} 
\Delta_{ij}(z) =
\left. \sum_X \int \frac{\de \xi^- }{2\pi \sqrt{2}} \,
e^{ik\cdot \xi} \langle 0 \vert \psi_i (\xi) \vert P_h, S_h, T_h;X\rangle
\langle P_h,S_h, T_h;X\vert\overline \psi_j(0) \vert 0 \rangle
\right|_{\xi^+ = \xi_T = 0},
\end{equation} 
depending on the light-cone momentum fraction $z= P_h^-/k^-$, where $P_h$ is
the momentum of the outgoing hadron and $k$ the momentum of the fragmenting
quark. 

The analysis of the leading-order scattering matrix can be developed in
complete analogy to the distribution correlation function. 
The matrix $M= \Delta \gamma^-$ now represents
a hadron-quark decay matrix, which is again positive definite. Instead of
using the hadron spin density matrix, we should employ the hadron decay 
matrix, expressing it in terms of vector and tensor analyzing powers $A_{L},
A_{T}, A_{LL}$, etc.
The final result, however, would still have the form of Eq.~(\ref{e:emme}),
with the $z$-dependent fragmentation functions $D_1, G_1, D_{1LL}, H_1$ and
$H_{1LT}$ replacing the $x$-dependent distribution functions 
$f_1, g_1, f_{1LL}, h_1$ and $h_{1LT}$. Despite the differences in notation,
our fragmentation functions correspond to the ones of Ji~\cite{ji} (i.e. 
$D_{1LL}=\hat{b}_1$ 
and $H_{1LT}=\hat{h}_{\ovl{1}}$).
Note that in the context of a fragmentation process 
it is not justified to
discard T-odd functions. 
We point out that the description in terms of the matrix in spin space allows a
connection to the analysis performed in~\cite{Anselmino:1996vq}.

It is particularly useful to set bounds on the
function $H_{1LT}$, which is a possible candidate to probe the transversity
distribution of the nucleon in inclusive deep-inelastic vector meson 
production. It appears in connection with the 
transversity distribution in a 
$\langle\sin{(\phi_{\pi \pi}^{\ell} + \phi_S^{\ell})}\rangle$ 
transverse 
spin asymmetry, where 
$\phi_S^{\ell}$ and  $\phi_{\pi \pi}^{\ell}$ are the azimuthal
angles with respect to the lepton scattering plane of, respectively, 
 the spin of the target 
and the difference of the momenta of the two decay particles~\cite{brook}.
   The equivalent of Eq.~(\ref{e:soffer}) for
fragmentation functions 
implies the bounds 
\begin{eqnarray}   
\left[H_{1LT}(z)\right]^2 + \left[H_1(z)\right]^2 &\leq& 
	\frac{1}{2}\Big[D_1(z)+\frac{2}{3}\,D_{1LL}(z)\Big] 
	\left[D_1(z)+G_1(z)-\frac{D_{1LL}(z)}{3}\right]  \nn \\
	&\leq&	\left[D_1(z)+\frac{2}{3}\,D_{1LL}(z)\right] 
		\left[D_1(z)-\frac{1}{3}\, D_{1LL}(z)\right] \; \leq \;
	\frac{9}{8}\, \left[D_1(z)\right]^2 .
\end{eqnarray} 
The less restrictive bounds are particularly relevant because 
in the specific case vector meson decay the vector 
analyzing powers are 
zero, making it impossible to measure the fragmentation functions $H_1$ and
$G_1$. Note that some information on the function  $D_{1LL}$ can be already
extracted from experimental measurements~\cite{Ackerstaff:1997kj,delphi}.

\section{Bounds on transverse momentum dependent functions}

To include the dependence on the parton transverse momentum, ${\bm p}_T$, 
the correlation function can be defined as~\cite{multa}
\begin{equation} 
\Phi_{ij}(x,{\bm p}_T) = 
\left. \int \frac{\de \xi^-\,\de {\bm \xi}_T}{(2\pi)^3}\ \e^{\ii p\cdot \xi}
\,\langle P,S,T\vert \overline \psi_j(0) \psi_i(\xi)
\vert P,S,T\rangle \right|_{\xi^+ = 0},
\end{equation}
For convenience, the correlation function can be decomposed in
several terms in relation to the polarization state of the target, {\it i.e.}\
$ 
\Phi = \Phi_U + \Phi_L +\Phi_T + \Phi_{LL} +\Phi_{LT} + \Phi_{TT}$. 
To leading order in $1/Q$,  these terms can be decomposed 
 as
(we identify T-odd terms by enclosing them in round parentheses)
\begin{eqnarray} 
\Phi_U (x, {\bm p}_T)& =& \frac{1}{4}\left\{
	f_1(x, p_T^2)\,\nslash_+ +
  \left(h_1^{\perp}(x, p_T^2)\,
		\sig_{\mu \nu}\frac{p_T^{\mu}}{M}  n_+^{\nu}\right) \right\},\\
\Phi_L (x, {\bm p}_T)& =& \frac{1}{4}\left\{
	g_{1L}(x, p_T^2)\,S_L \,\g_5 \,\nslash_+ +
	h_{1L}^{\perp}(x, p_T^2)\,S_L\,
		\ii \sig_{\mu \nu}\g_5 n_+^{\mu}\frac{ p_T^{\nu}}{M}\right\},\\
\Phi_T (x, {\bm p}_T)& =& \frac{1}{4}\left\{
	g_{1T}(x, p_T^2)\,\frac{{\bm S}_T\cdot{\bm p}_T}{M}\,\g_5 \,\nslash_+ +
	h_{1T}(x, p_T^2)\,
		\ii \sig_{\mu \nu} \g_5 n_+^{\mu} S_T^{\nu} \right. \nn \\
&&\left.\mbox{}+
	h_{1T}^{\perp}(x, p_T^2)\,\frac{{\bm S}_T \cdot {\bm p}_T}{M}\, 
	     \ii \sig_{\mu \nu}\g_5 n_+^{\mu}\frac{p_T^{\nu}}{M} \right. \nn \\
&&\left.\mbox{}+
  \left(f^{\perp}_{1T}(x, p_T^2)\,\eps_{\mu \nu \rho \sig}
	   \g^{\mu}n_+^{\nu}\frac{p_T^{\rho}}{M}S_T^{\sig} \right)\right\}, \\
\Phi_{LL} (x, {\bm p}_T)& =& \frac{1}{4}\left\{
	f_{1LL}(x, p_T^2)\,S_{LL} \,\nslash_+ +
  \left(h_{1LL}^{\perp}(x, p_T^2)\,S_{LL} \,
		\sig_{\mu \nu} \frac{p_T^{\mu}}{M}n_+^{\nu}\right) \right\}, \\
\Phi_{LT} (x, {\bm p}_T)& =& \frac{1}{4}\left\{
	f_{1LT}(x, p_T^2)\,\frac{{\bm S}_{LT} \cdot {\bm p}_T}{M}\, \nslash_+ +
  \left(g_{1LT}(x, p_T^2)\,\eps_T^{\mu \nu}S_{LT\,\mu}\frac{p_{T\,\nu}}{M} 
		\,\g_5 \,\nslash_+\right) 	\right. \nn \\ 
&&\left.\mbox{}+
  \left(h'_{1LT}(x, p_T^2)
    \, \ii \sig_{\mu \nu} \g_5 n_+^{\mu}\eps_T^{\nu \rho}S_{LT\,\rho}\right)
			\right. \nn \\ 
&&\left.\mbox{}+
  \left(\,h_{1LT}^{\perp}(x, p_T^2)\, \frac{{\bm S}_{LT} \cdot {\bm p}_T}{M}\,
			\sig_{\mu \nu}\frac{p_T^{\mu}}{M} n_+^{\nu} 
			\right) \right\},\\
\Phi_{TT} (x, {\bm p}_T)& =& \frac{1}{4}\left\{
   f_{1TT}(x, p_T^2)\,\frac{{\bm p}_T \cdot {\bm S}_{TT} \cdot {\bm p}_T}{M^2}
		\, \nslash_+  \right. \nn \\ 
&&\left.\mbox{}-
  \left(g_{1TT}(x, p_T^2)\, \eps_T^{\mu \nu} S_{TT\,\nu \rho}
	\frac{p_T^{\rho} p_{T\,\mu}}{M^2}\,\g_5\,\nslash_+\right)\right. \nn \\
&&\left.\mbox{}-
  \left(h'_{1TT}(x, p_T^2)\, \ii \sig_{\mu \nu} \g_5 n_+^{\mu}
			\eps_T^{\nu \rho} S_{TT\,\rho \sig}
	\frac{p_T^{\sig}}{M} \right)
			\right. \nn \\ 
&&\left.\mbox{}+
  \left(\,h_{1TT}^{\perp}(x, p_T^2)\,
		\frac{{\bm p}_T \cdot {\bm S}_{TT} \cdot {\bm p}_T}{M^2}\,
		\sig_{\mu \nu} \frac{p_T^{\mu}}{M} n_+^{\nu} \right) \right\}.
\end{eqnarray} 

Following steps analogous to the previous section, we can reconstruct the
complete $6 \times 6$ scattering matrix. The inclusion of ${\bm p}_T$ 
dependence makes all the entries of the
matrix to be non-zero. It is also convenient to define the functions
\begin{eqnarray} 
h_{1LT}(x, {p}^2_T)
       &=& h'_{1LT} (x,{p}^2_T)
	+ \frac{{\bm p}_T^2}{2M^2}\;h_{1LT}^{\perp}(x,{p}^2_T), \\
h_{1TT}(x,{p}^2_T) 
	&=& h'_{1TT} (x,{p}^2_T)
	+ \frac{{\bm p}_T^2}{2M^2}\;h_{1TT}^{\perp}(x,{p}^2_T).
\end{eqnarray} 
In the rest of this Section, unless
otherwise specified, all the functions are understood to depend on the 
variables $x$ and $p_T^2$. 

In principle, if we fully exploit the condition of the scattering matrix to be
positive definite, we can write several relations involving an
increasing number of different functions. We feel this to be an excessive task
if compared to the exiguity of information we have on the involved functions.
Therefore, here 
we choose to focus on the relations stemming from positivity of the 
two-dimensional minors of the matrix.

Because of the symmetry properties of the matrix, 
only nine
independent inequality relations between the different functions are produced.
To simplify
the discussion, it is useful to identify some of the T-odd functions as 
imaginary parts
of the T-even functions, which become then complex scalar functions. The
following replacements are required 
\begin{eqnarray} 
g_{1T}-\ii \, f_{1T}^{\perp} &\rightarrow& g_{1T}, \\
f_{1LT}-\ii \, g_{1LT} &\rightarrow& f_{1LT} , \\
f_{1TT}-\ii \, g_{1TT}^{\perp} &\rightarrow& f_{1TT}, \\
 h_{1}+\ii \, h_{1LT} &\rightarrow& h_{1}, \\
h_{1T}^{\perp}-\ii  \,h_{1LT}^{\perp} &\rightarrow& h_{1T}^{\perp}.
\end{eqnarray} 

Assuming the above relations, we can write the following inequalities:
\begin{eqnarray} 
\vert h_{1} \vert^2 & \leq & 
	\frac{1}{2}\,\left(f_1+\frac{2\,f_{1LL}}{3}\right) 
		\left(f_1+g_1-\frac{f_{1LL}}{3}\right) ,  \\
\frac{\vert p_{T}\vert^2}{2 M^2} \; \vert g_{1T} + f_{1LT} \vert^2 & \leq &
	\left(f_1+\frac{2\,f_{1LL}}{3}\right) 
		\left(f_1+g_1-\frac{f_{1LL}}{3}\right) ,  \\
\frac{\vert p_{T}\vert^2}{2 M^2} \; \vert g_{1T} - f_{1LT} \vert^2 & \leq &
	\left(f_1+\frac{2\,f_{1LL}}{3}\right) 
		\left(f_1-g_1-\frac{f_{1LL}}{3}\right) ,  \\
\frac{\vert p_{T}\vert^4}{2 M^4} \; \vert h_{1T}^{\perp}\vert^2 & \leq &
	\left(f_1+\frac{2\,f_{1LL}}{3}\right) 
		\left(f_1-g_1-\frac{f_{1LL}}{3}\right) ,  \\
\frac{\vert p_{T}\vert^6}{M^6} \;{h_{1TT}^{\perp}}^2 & \leq & 
	\left(f_1-g_1-\frac{f_{1LL}}{3}\right)^2 ,  \\
\frac{\vert p_{T}\vert^2}{4 M^2} 
	\;\left(h_{1}^{\perp} +\frac{2\, h_{1LL}^{\perp}}{3}\right)^2 & \leq & 
	\left(f_1+\frac{2\,f_{1LL}}{3}\right)^2 ,  \\
\frac{\vert p_{T}\vert^2}{M^2} \;h_{1TT}^2 & \leq & 
	\left(f_1+g_1-\frac{f_{1LL}}{3}\right)^2 ,  \\
\frac{\vert p_{T}\vert^2}{M^2} \; \left[{h_{1L}^{\perp}}^2 +
	\left(h_{1}^{\perp} -\frac{h_{1LL}^{\perp}}{3}\right)^2\right] & \leq &
	\left(f_1+g_1-\frac{f_{1LL}}{3}\right)
	\left(f_1-g_1-\frac{f_{1LL}}{3}\right), \\	
\frac{\vert p_{T}\vert^4}{4 M^4} \; \vert f_{1TT} \vert^2 & \leq &
	\left(f_1+g_1-\frac{f_{1LL}}{3}\right)
	\left(f_1-g_1-\frac{f_{1LL}}{3}\right).
\end{eqnarray} 

A further simplification can be performed by introducing the positive real
 functions
\begin{eqnarray} 
A(x,p_T^2)&\equiv& f_1(x,p_T^2)+g_1(x,p_T^2)-\frac{1}{3}f_{1LL}(x,p_T^2),  \\
B(x,p_T^2)&\equiv& f_1(x,p_T^2)+\frac{2}{3}\, f_{1LL}(x,p_T^2),  \\
C(x,p_T^2)&\equiv& f_1(x,p_T^2)-g_1(x,p_T^2)-\frac{1}{3}f_{1LL}(x,p_T^2)  
\end{eqnarray} 
and express the combinations of distribution functions occurring in the
various entries of the matrix as
\begin{equation} 
\begin{array}{rclrcl}
h_{1} & = & a\,\sqrt{\frac{1}{2}\,A\,B} ,\qquad &
	h_{1}^{\perp} +\frac{2}{3}\, h_{1LL}^{\perp} &=& f\,B , \\
g_{1T} + f_{1LT} &=& b\,\sqrt{A\,B} , &
	h_{1TT} &=& g\,A ,\\
g_{1T} - f_{1LT} &=& c\,\sqrt{C\,B} , &
 h_{1L}^{\perp} - \ii \left(h_{1}^{\perp} -\frac{1}{3}\,h_{1LL}^{\perp}\right) 
	&=&  h\,\sqrt{A\,C} ,\\
h_{1T}^{\perp}&=& d\,\sqrt{C\,B} ,  &
	f_{1TT} &=& j\,\sqrt{A\,C}, \\
h_{1TT}^{\perp} &=& e\,C , 
\end{array}
\end{equation}
where the coefficients $a$, $b$, $c$, $d$, $h$, $j$ are complex functions of
the variables $x$ and $p_T^2$, and the coefficients $e$, $f$, $g$ are real
functions of the same variables. Time-reversal invariance in the distribution
sector constrains the coefficients $a$, $b$, $c$, $d$, $h$, $j$ to be real
functions of the variables $x$ and $p_T^2$ and the coefficients $e$, $f$, $g$
to vanish.

The scattering matrix can now be expressed in a concise way as
\begin{eqnarray} 
\lefteqn{M_{i \lambda,\,j \lambda'}(x,{\bm p}_T) =} \nn \\
&&\left\lgroup \begin{array}{cccccc}
	A & 
	\e^{-\ii \phi}\,b\,\sqrt{A\,B} & 
	\e^{-2 \ii \phi}\,j\,\sqrt{A\,C} & 
	\e^{\ii \phi}\,h\,\sqrt{A\,C} &
	a\,\sqrt{A\,B}& 
	- \ii \e^{-\ii \phi}\,g\,A \\
& \\	
	\e^{\ii \phi}\,b^{\ast}\,\sqrt{A\,B} & 
	B & 
	\e^{-\ii \phi}\,c\,\sqrt{C\,B} &
	\e^{2 \ii \phi}\,d\,\sqrt{C\,B}& 
	- \ii \e^{\ii \phi}\,f\,B&
	a^{\ast}\,\sqrt{A\,B} \\
& \\
	\e^{2 \ii \phi}\,j^{\ast}\,\sqrt{A\,C} & 
	\e^{\ii \phi}\,c^{\ast}\,\sqrt{C\,B}&
	C&
	-\ii \e^{3 \ii \phi}\,e\,C &
	\e^{2 \ii \phi}\,d^{\ast}\,\sqrt{C\,B}&
	-\e^{\ii \phi}\,h^{\ast}\,\sqrt{A\,C} \\
& \\
	\e^{-\ii \phi}\,h^{\ast}\,\sqrt{A\,C}  & 
	\e^{-2 \ii \phi}\,d^{\ast}\,\sqrt{C\,B}&
	\ii \e^{-3 \ii \phi}\,e\,C&
	C&
	-\e^{-\ii \phi}\,c^{\ast}\,\sqrt{C\,B}&
	\e^{-2 \ii \phi}\,j^{\ast}\,\sqrt{A\,C}\\
& \\
	a^{\ast}\,\sqrt{A\,B} & 
	\ii \e^{-\ii \phi}\,f\,B&
	\e^{-2 \ii \phi}\,d\,\sqrt{C\,B}&
	-\e^{\ii \phi}\,c\,\sqrt{C\,B}&
	B&
	-\e^{\ii \phi}\,b^{\ast}\,\sqrt{A\,B}\\
& \\
	 \ii \e^{\ii \phi}\,g\,A  & 
	a\,\sqrt{A\,B}&
	-\e^{-\ii \phi}\,h\,\sqrt{A\,C}&
	\e^{2 \ii \phi}\,j\,\sqrt{A\,C}&
	-\e^{-\ii \phi}\,b\,\sqrt{A\,B}&
	A
\end{array}\right\rgroup.\nn \\ 
\end{eqnarray}
Positivity of the two-dimensional minors of this matrix requires the absolute 
value  of the 
coefficients $a$ to $j$ to be smaller or equal to unity. It is possible to
consider higher dimensional minors to extract other
useful bounds. The choice of the appropriate minors will depend on the 
availability of measurements of specific functions.

\section{Conclusions}

In this letter, we studied the leading-order part of the correlation functions
for parton
distribution inside spin-one targets or parton fragmentation into spin-one 
hadrons. We cast the correlation functions in the form
of scattering matrices in the parton chirality
space $\otimes$ hadron spin space. Positive definiteness of these matrices 
allowed us to set bounds to the involved
distribution and fragmentation functions.
These elementary bounds can act as important guidances to estimate the
magnitude of otherwise unknown functions and, consequently, the magnitude of
spin and azimuthal asymmetries appearing in several processes containing
spin-one hadrons. In particular, these bounds are significant for
one-particle-inclusive deep inelastic scattering and $e^+ e^-$ annihilation
with vector mesons in the final state.
Among our results, we presented a bound that generalizes the Soffer 
inequality to the
case of spin-one targets.
Furthermore, we proposed a bound on the fragmentation function
$H_{1LT}$. Since the latter can appear in connection with the 
transversity distribution in a 
$\langle\sin{(\phi_{\pi \pi}^{\ell} + \phi_S^{\ell})}\rangle$ 
 asymmetry, our bound can be useful in the 
extraction of the transversity distribution, where, for instance,
 it could be important to constrain the absolute normalization of the
function.

\acknowledgements{This work is supported by the Foundation for Fundamental
Research on Matter (FOM) and the Dutch Organization for Scientific Research 
(NWO).}


\end{document}